\documentclass[a4paper,12pt]{article}
\usepackage{eurosym}
\usepackage[center]{titlesec}
\usepackage{lipsum}
\usepackage[utf8]{inputenc}
\usepackage{cancel}
\usepackage{tikz}
\usepackage{ulem}
\usepackage{amsfonts}
\usepackage{amssymb}
\usepackage{graphicx}
\usepackage{amsmath}
\usepackage{enumerate}
\usepackage{mathtools}
\usepackage{subfig}
\usepackage{color}
\usepackage{tikz}
\usepackage{float}
\usepackage{here}
\usepackage{cite}
\usepackage{mathrsfs}
\usepackage{float,epsfig}
\usepackage{dcolumn}
\usepackage{graphicx}
\usepackage{bm}
\usepackage{amsmath,amssymb,amsthm}
\usepackage[colorlinks=true,linkcolor=blue,citecolor=red]{hyperref}
\usepackage{multirow}
\usepackage[toc,page]{appendix}

\renewcommand{\thesection}{\Roman{section}}
\renewcommand{\thesubsection}{\arabic{subsection}}
\renewcommand{\thesubsubsection}{\alph{subsubsection}}
\titleformat
    {\section}
    [block]
    {\centering\bfseries}
    {\thesection.}
    {0.5em}
    {}
\titleformat
    {\subsection}
    [block]
    {\centering\itshape}
    {\thesubsection.}
    {0.5em}
    {}
\titleformat
    {\subsubsection}
    [block]
    {\centering}
    {\thesubsubsection.)}
    {0.5em}
    {}
\titlespacing{\section}{0pt}{*4}{*1.5}
\titlespacing{\subsection}{0pt}{*4}{*1.5}
\titlespacing{\subsubsection}{0pt}{*4}{*1.5}
\usetikzlibrary{arrows.meta}
\usetikzlibrary{bending}
\usetikzlibrary{calc}
\newcommand{\be}{\begin{equation}}
\newcommand{\ee}{\end{equation}}
\newcommand{\bea}{\setlength\arraycolsep{2pt} \begin{eqnarray}}
\newcommand{\eea}{\end{eqnarray}}

\def\0{{\sst{(0)}}}
\def\1{{\sst{(1)}}}
\def\2{{\sst{(2)}}}
\def\3{{\sst{(3)}}}
\def\4{{\sst{(4)}}}
\def\5{{\sst{(5)}}}
\def\6{{\sst{(6)}}}
\def\7{{\sst{(7)}}}
\def\8{{\sst{(8)}}}
\def\sst#1{{\scriptscriptstyle #1}}

\makeatletter \@addtoreset{equation}{section}

\definecolor{lime}{HTML}{A6CE39}

\begin{document}

\title{ {\normalsize \textbf{\Large On Thermodynamics of   Charged Black Holes,
Swampland, and Dark Matter }}}
\author{ {\small Saad Eddine Baddis$^{1}$,  Adil Belhaj$^{1}$, Hajar Belmahi$^{1,2}$,  Salah Eddine Ennadifi$^{3}$\thanks{Authors are listed in alphabetical order. } \thanks{Corresponding author:   saadeddine.baddis@um5r.ac.ma. } \hspace*{-8pt}} \\
{\small $^{1}$ESMaR, Faculty of Science, Mohammed V University in Rabat, Rabat,
Morocco} \\
{\small $^{2}$National School of Applied Sciences (ENSA), Chouaib Doukkali University, El Jadida, Morocco} \\
{\small  $^{3}$LHEP-MS, Faculty of Science, Mohammed V University in Rabat, Rabat, Morocco}}
\maketitle

\begin{abstract}
{\noindent } Inspired by the idea that the cosmological constant can be
considered as a dynamical  quantity, we present a scenario bridging
certain swampland conjectures from a new look at thermodynamics   of  black holes.
Dealing with  the radial metric function at the horizon as an equation of state, we
discuss   structural transitions and coexistence curves. By considering an effective 
potential with a quasi-static scalar field that could find a place in string theory, we then establish a connection between certain swampland conjectures.  Relying on Kaluza Klein interpretations of scalar fields, we approach  the dark dimension and dark matter
through such a  thermodynamic approach to charged black holes.

\end{abstract}

%

\newpage


\newpage

\section{Introduction}

Over the past decade, the thermodynamics of black holes has attracted
considerable interest within the context of various theories of gravity. It
has been the subject of studies supported by empirical observations, notably
the acquisition of the first image of certain massive black holes by the
international Event Horizon Telescope (EHT) collaboration\cite%
{f10,f11,f12,f14}. These developments have been made possible mainly
thanks to the link between the cosmological constant and the pressure \cite%
{T0, T1}. This link has led to the derivation of many interesting results in
thermodynamic systems incorporated into non-trivial theories of gravity.
Following this correspondence, various structure transitions have been
studied, yielding certain universal relationships that serve as powerful
tools for testing models of gravity. In particular, the thermodynamic
transitions and the criticality behaviors  of anti de Sitter (AdS) black
holes have been extensively investigated using different approaches \cite%
{T2,T3,T4,T5}. More precisely, an interplay between the AdS black hole phases
and the statistical configurations corresponding to Van der Waals like phase
transitions has been established by considering the cosmological constant as
a thermodynamical quantity, and its conjugate variable as the thermodynamic
volume \cite{T1,T2}. These techniques have been extended to non-trivial
theories such as type II superstrings and M-theory involving different brane configurations. In this way, the cosmological constant has been related to the
brane number offering a way to examine the thermodynamic stability of the  AdS
black holes in higher dimensional spacetimes \cite{ST1,ST2}. -

Recently, the swampland criteria program has received a remarkable interest
in connection with various supergravity theories associated with high energy
physics including superstring models and black holes \cite{SL1,SL2,SL3,SL4,SL5,SL6}. This program appears to have a significant
influence on how effective field theories (EFTs) are approached and their
compatibility with quantum gravity. Effectively, this allows one to
distinguish EFTs that are consistent with quantum gravity from those that
are incompatible with such a coupling. This approach draws inspiration from
non-trivial theories of gravity that incorporate black holes and string
theory. Furthermore, it has been suggested that the swampland criteria have
also been developed in relation with the dark dimension (DD) and the dark
matter (DM) \cite{D1,D2,D3}. More recently, an emphasis has been placed in
particular on their application to the inflationary scenario \cite{RG1,RG2,RG3,RG4,RG5,RG6,BB1,BB2}. In such investigation activities, various conjectures
have been extensively assessed using different angles of thinking. The most dealt with
are the weak gravity conjecture (WGC) and the distance conjecture (DC). The first one 
postulates that a theory involving a gauge field coupled to gravity should
include charged matter for which gravity has the weakest effect \cite{WGC1,WGC2,WGC3,WGC5,WGC6,WGC7,WGC8,WGC9}. However, the second states
that, in any theory of quantum gravity, an infinite shift in the
scalar field moduli space leads to the emergence of a tower of
particles whose masses become exponentially very small \cite{DC1,DC2,DC3,DC4,DC5,DC6,DC7,DC8,DC9}. At low energy, this indicates that
EFTs collapse when fields travel over large distances.

Inspired  by  a dynamical model of the cosmological constant,  we propose a scenario  which 
bridges these two conjectures, through the thermodynamics of black holes.
More specifically, we propose a new way of looking at the thermodynamics of
such objects  by paying attention to the metric function, which holds
information on the underlying physical properties. Interpreting such a
function at the horizon  as  a state equation,  we discuss  phase transitions and coexistence curves.  By considering an effective potential
associated with a quasi-static scalar field that may play a role in string theory, we then establish a connection  between certain swampland conjectures. Using Kaluza-Klein interpretations of scalar fields, we  approach  the 
DD  and DM
through such a thermodynamic view of charged black holes.

The organization of this work is as follows. In section 2, we provide a new
take on the black holes thermodynamics. Section 3 is devoted to the
coexistence curve in black hole physics. In section 4, we establish an interplay
between certain swampland conjectures with DM  implications. The
last section gives concluding remarks.

\section{On new insights into the thermodynamics of black holes}

The cosmological constant has played a crucial role in the study of the
thermodynamic behavior of black holes, where various aspects have been
addressed, including phase transitions and P-V criticality \cite{T0,T1}. In
this approach, it has been identified with the pressure of black holes
considered as thermodynamic objects. Inspired by the fact that under
the consideration of  the microscopic spacetime fluctuations at
Planckian scales  ($M_{Planck}$),  the cosmological constant can
be regarded as an effective quantity $\Lambda _{eff}=\Lambda (t)$
reflecting the backreaction of such fluctuations on the large-scale geometry
\cite{CC}, we propose a new approach to the thermodynamics of black holes by
focusing on the metric function, which contains information about the
underlying physical properties. A priori, there could be many approaches to
introduce the dynamical cosmological  constant into black hole thermodynamics. This
undoubtedly entails certain modifications to the Einstein equations of motion
by altering the associated energy-momentum tensor. One possible approach is
to introduce an effective potential with a quasi static scalar field. This
is a time-dependent field where the evolution is slow enough that its
spatial structure, over a given time interval, is similar to a static field configuration. In this case, the time derivatives can be neglected in the corresponding equations of motion. This type of situation has been encountered
most often in cosmology and modified gravity, where the calculations are
simplified by assuming that the scalar field can be dealt with as a
perturbation of matter without high-frequency oscillations. Through this view, roughly, we establish a connection with the swampland program and
related topics including DD and DM. Precisely, we show that certain
swampland conjectures can be bridged by considering an alternative take on
charged black hole phase transitions. To start, we consider black hole
configurations which can be derived from the following action 
\begin{equation}
S=\int d^{4}x\sqrt{-g}\left[ \frac{R}{2\kappa ^{2}}-\frac{1}{4}F^{\mu \nu
}F_{\mu \nu }-\frac{1}{2}\partial _{\mu }\phi \partial ^{\mu }\phi -V(\phi )%
\right]   \label{e0021}
\end{equation}%
where one has used  $F_{\mu \nu }=\partial _{\mu }A_{\nu }-\partial _{\nu
}A_{\mu }$ where $A_{\mu }$ is the Maxwell gauge vector. $\phi $  represents  a real
scalar field with a potential $V(\phi )$. Up to an appropriate  approximations, the
Einstein field equations could provide a black hole metric function
describing (static) symmetric solutions via the following  line element form 
\begin{equation*}
ds^{2}=f(r)dt^{2}-\frac{1}{f(r)}dr^{2}+r^{2}d\Omega ^{2}
\end{equation*}%
where $d\Omega ^{2}$ represents the metric on the two dimensional unit
sphere, usually denoted by $S^{2}$. The above action Eq.(\ref{e0021})
provides the following Einstein field equations 
\begin{equation}
R_{\mu \nu }-\frac{1}{2}g_{\mu \nu }R=T_{\mu \nu },
\end{equation}%
where the tensor $T_{\mu \nu }$ containing matter contributions can be  expressed  as 
\begin{equation}
T_{\mu \nu }=\partial _{\mu }\phi \partial _{\nu }\phi -\frac{1}{2}g_{\mu
\nu }\partial _{\sigma }\phi \partial ^{\sigma }\phi +F_{\mu \sigma }F_{\nu
}^{\sigma }-\frac{1}{4}g_{\mu \nu }F_{\rho \sigma }F^{\rho \sigma }-g_{\mu
\nu }\frac{V(\phi )}{k^{2}}.
\end{equation}%
To establish the interplay that we are  after, certain requirements should be
imposed on the scalar scalar field to provide dynamical
contributions similar to cosmological ones. Considering the following constraints 
\begin{equation*}
\phi \equiv \phi (t)\qquad \partial _{t}\phi \partial ^{t}\phi \approx 0
\end{equation*}%
being compatible with the slow-roll conditions $(\dot{\phi}^{2}<<(V(\phi
)\leq \mathcal{O}(1))$, the assumed ansatz of the black hole line element
can take the following form

\begin{equation}
f(r)=1-\frac{2M}{r}+\frac{Q^{2}}{r^{2}}-V(\phi )r^{2}
\end{equation}%
where the unit systems have been used. $M$ and $Q$ are the mass and the
electric charge of the associated the black hole, respectively. To assess the associated
thermodynamics, the scalar potential form should be precised. Forgetting a
while about such a detail, we focus on the metric function $f(r)$. Motivated
by many thermodynamic features, we alternatively approach such a function
carrying data on black hole properties including the optical ones. Indeed,
this function can be interpreted as a state equation by imposing the
constraint 
\begin{equation}
f(r)=0.  \label{1}
\end{equation}%
With this insight, we are building similarities between  the
thermodynamics variables and some  black hole physics quantities. In such manner, the effective scalar potential $V(\phi )$ can play the role of the
pressure which is usually linked to the cosmological constant. In the
present context, this quantity becomes an effective pressure taking the
following form 
\begin{equation}
P=-\frac{3V(\phi )}{8\pi }  \label{100}
\end{equation}%
which may be seen as as an extension of the one related to the cosmological
constant providing the ordinary thermodynamic approach of charged black
holes. A close examination shows that the temperature can be expressed in
terms of the scalar field $\phi$ and the  electric charge  $Q$
\begin{equation*}
T=T(Q,\phi ).
\end{equation*}%
As in the investigation of criticality behaviors, the volume is replaced by
the outer horizon $r^{+}$ being a real solution of $f(r)=0$. With
these similarities at hand, Eq.(\ref{1}) can be viewed as a state equation in
black hole physics which reads as follows 
\begin{equation}
f(r^{+},Q,\phi )=0.  \label{2}
\end{equation}%
In this way, we thus replace $P$, $T$ and $V$ by $\phi $, $Q$ and $r^{+}$.
Before going ahead, the scalar potential form should be precised. A priori, there
are of course many models being investigated in the context of non-trivial
gravities including string theory and related topics.  Precisely, the 
scalar field spectrums can be derived using different mechanisms such as
either the Calabi-Yau compactification or the KK  mechanism where the 
scalar fields  such as the  dilaton or  the moduli naturally appear
in the low-energy effective action. Effectively, they   are typically associated with
exponential couplings \cite{K1,K2,K3,K4,K5}. In  this way, the dimensional reduction
from higher dimensions leads to exponential form potentials $V(\phi )\sim
Ce^{-k\phi }$\ where the shape or the  size of extra dimensions  could  
provide  the scalar field $\phi$   and  $C$  is a model-dependent
constant.  $k$ can be viewed as a parameter weighing the
coupling strength of the scalar field. In the present model, and to ensure a
positive thermodynamic pressure \cite{100},  we are considering, broadly speaking, the
scalar potential
\begin{equation}
V(\phi )=-\frac{e^{-2a\phi }}{\ell _{p}^{2}}  \label{po}
\end{equation}%
where $\ell _{p}$  represents  the Planck length and $a$ 
is a free parameter.  

\section{On coexistence curve in black hole physics}

In this section, we investigate the small/large black hole transition using
such an insight on the associated thermodynamics. It is recalled that this
transition can be approached through the Gibbs free energy being a
thermodynamic potential playing a crucial role in describing and predicting
the corresponding  physical behaviors 
\begin{equation}  \label{dG}
dG=-SdT+\Phi dQ+ \mathcal{V}dP
\end{equation}
where $S$ represents the entropy of the system and $\Phi $ is the electric
scalar  potential given by 
\begin{equation}
\Phi= \frac{Q}{r^+}.
\end{equation}
$P$ and $\mathcal{V}$ denote the thermodynamic pressure and the  volume, 
respectively. In the present work, we reconsider the phase transitions in
the context of light and heavy states supported by string theory features. To
do so, we set certain phase transition criteria for small $(s) $ and large
black hole $(\ell)$.

\begin{itemize}
\item The mechanic and electric equilibrium between two such states are
required by 
\begin{equation}
\phi_s=\phi_{\ell }, \qquad Q_s=Q_{\ell}.
\end{equation}

\item The continuity of the Gibbs free energy between the two phases is
ensured by 
\begin{equation}
G_{s}=G_{\ell}.
\end{equation}

\item The discontinuity of the first partial derivative of the free energy
with respect to the extensive variables involved  is  conditioned by 
\begin{equation}
\frac{\partial G_s}{\partial Q}\neq\frac{\partial G_{\ell}}{\partial Q},
\qquad \frac{\partial G_s}{\partial \phi}\neq\frac{\partial G_{\ell}}{%
\partial \phi}.\label{Connection}
\end{equation}
\end{itemize}

It is worth mentioning that the electrical equilibrium actually follows from the assumption that the phase transition is expected to respect a global symmetry, by requiring a gauge-invariant phase for the associated operators. In fact, the symmetry argument presented here is analogous to the treatment of the latent heat in classical thermodynamics. In the same way that the  latent heat contributes to phase transformations at constant temperature, the symmetry argument requires a potential gradient contributing to the phase change with a conserved charge. This last point is entirely justified  one  considers that the Hawking evaporation is insensitive to global symmetries. According to \cite{ GS1,GS2}, the main argument of this work  is that the lifetime of black holes is bounded by two phases, for which the Hawking-Bekinstein formulation works well for one but fails for the other.  In what follows,  we provide a  discussion  dealing with this issue.

\begin{figure}
\begin{center}
\begin{tabbing}
\includegraphics[scale=0.55]{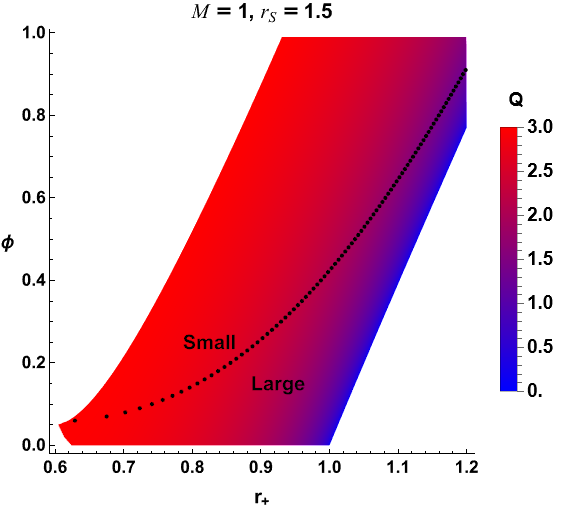}
\hspace{0.2cm} \includegraphics[scale=0.55]{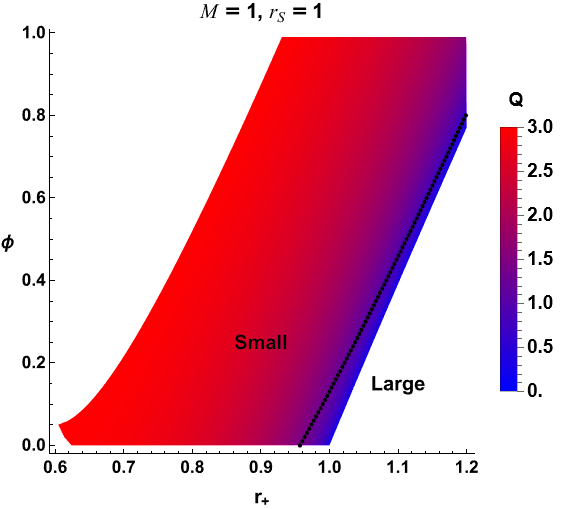}\\
  \end{tabbing}
\caption{ $(\protect\phi -r_{+})$-diagrams in the AdS case for different small phase sizes $r_s.$}
\label{Fig1}
\end{center}
\end{figure}

Following the works reported in \cite{ Thermo1,Thermo2 }, one can show that the coexistence curve can be approached from
Clausius-Clapeyron equations. A priori, there are certain relations which could be exploited. However,
here, we consider the esothermal generalized Clausius-Clapeyron equation 
\begin{eqnarray}  \label{TG}
dG_\ell)_T-dG_s)_T=0.
\end{eqnarray}
At fixed temperature values, Eq.(\ref{dG}) and Eq.(\ref{TG}) provide the following
differential equation 
\begin{equation}
Q\frac{dQ}{d\phi}=\frac{3a}{8\pi \ell^2_p}\frac{\mathcal{V}_\ell-\mathcal{V}%
_s}{r^+_s-r^+_\ell}r_s^+r^+_\ell e^{-2a\phi}
\end{equation}
Using the volume relations $\mathcal{V}_\ell= \frac{4\pi (r^+_\ell)^3 }{3} $
and $\mathcal{V}_\ell= \frac{4\pi (r^+_s )^3}{3} $, this equation can be
expressed as 
\begin{equation}
Q\frac{dQ}{d\phi}=\frac{a}{ 2 \ell^2_p} \left( (r^+_\ell- r^+_s)^2+
3r^+_sr^+_\ell \right) r_s^+r^+_\ell e^{-2a\phi}
\end{equation}
To handle this differential equation in a nice way, we use the following identification 
limit

\begin{eqnarray}
r^+_\ell&=& \frac{3+\sqrt{5}}{ 2 }r_s.
\end{eqnarray}
In this way, the differential equation reduces to 
\begin{equation}
Q\frac{dQ}{d\phi}=-\frac{2a}{ \ell^2_p} r_s^4e^{-2a\phi}
\end{equation}
Integrating  out this equation, we get the following charge solution 
\begin{equation}  \label{coex}
Q=\frac{\sqrt{2}}{ \ell_p} r_s^2e^{-a\phi}
\end{equation}
interpreted as a coexistence curve separating large black hole phases and
small ones. Equipped with GPU accelerated CUDA computing techniques and in order to  illustrate the present proposition,  we consider the 
$(\phi -r_{+})$-diagram playing a similar role as the $(P-V)$ curve\footnote{It is recalled that CUDA is considered a flexible  computing platform and a set of programming models that exploit the parallel computing capabilities of NVIDIA GPUs  \cite{CUDA1,CUDA2,CUDA3,CUDA4,CUDABH1,CUDABH2,CUDABH3}. In this  way, the GPU architecture allows numerical computations  to be distributed across a large number of simultaneous multiprocessors (SMs), enabling highly efficient parallel processing and  programing.}.  Roughly,  Fig.(\ref{Fig1}) represents such  computational behaviors.

For fixed field values, it has been observed that the small black hole
appear for large charge values. However, large black holes are associated with small charge values. Additionally, we observe that as the size parameter $r_s$ increases, the large phases ability to store charge also increases. On the other hand, small phases become accessible for larger distances. It is worth noting that the presented formulation of the phase transition can be extended beyond the shape of  black hole event horizons encoded in the metric functions and even to the cases without hair. However, it is a tough task to develop  the associated  equation of state which needs more  thinking.

\newpage
\section{ Swampland, dark dimension, and dark matter}

We have established that charged  black holes can be distinguished by their two phases. However, as noted in \cite{Rem1,Rem2,Rem3,Rem4,Rem5}, the fact that the phenomenon of black hole evaporation does not take global symmetries into account raises the question of the existence of a large number of global residual charges, known as the  remnant problem.  In this section, the objective is to remedy the situation by classifying black hole phases according to their compliance with the Hawking-Bekinshtein formulation. More specifically, this will be useful in the reconsideration in the swampland program conjectures. Indeed, large phases are expected to strictly obey the above-mentioned formulation, whereas small phases do not necessarily conform to this description.
Based on the arguments given in \cite{GS2,SL2,SL6}, the breaking point of the Hawking-Bekingstein formulation is reached precisely when the measure around the black hole horizon hits a charged operator, which effectively leads to a naked singularity. Consequently, the conjecture regarding the absence of global symmetries  has been strongly supported in addressing this issue. Furthermore, using the formulation of \cite{GS3,GS4,GS5,GS6}, these works imply that global symmetries are merely local behaviors and must be gauged in such a way that the corresponding Lie group is generated by holonomy operators whose gauge phase is the measure of the connection associated with the Lie group around the black hole horizon, and a compact direction of size $r_s$.  Effectively,  this could  introduce  some sort of UV completion of the theory.

Making use of the argument that light states probe string
degrees of freedom, as they exist at the scalar level where QG effects are
accessible and where global symmetries are gauged or broken at high
energies. Furthermore, when the size of the black hole becomes compatible
with the size of the compact dimension (small black hole phase in the
present case), this saturates the  WGC  while remaining
consistent with the absence of  the global symmetry conjecture. The same
treatment can be applied to large black holes. More precisely, we argue that
they must exist uniquely at the IR scale, where the  local global symmetry behavior take effect, such that they obey the WGC by identifying with the inequality 
\begin{equation}
M^2-Q^2-9V(\phi)M^2Q^2+232V(\phi)Q^4-3712V(\phi)^2Q^6+52V(\phi)M^4\geq 0.
\end{equation}

For small mass and charge contributions, this reduces to the ratio
\begin{equation}
\frac{Q}{M}\leq 1.  \label{Eq1}
\end{equation}%
This construction enables the identification of light states with small
black hole phases. It is interesting to note that,  in terms of the  phase
size and the  moduli distance, Eq.(\ref{Eq1}) becomes 
\begin{equation}
r_{s}\leq R_{0}e^{a\phi /2}  \label{Eq2}
\end{equation}%
where the coexistence curve algebraic equation has been used with $%
R_{0}=2^{-1/4}\sqrt{M\ell _{p}}$. The graphical representation allows
a possible identification with the distance conjecture. We observe that the moduli
distance augments when the size of the light states is large. Given the
compatibility with the size of the compact dimension of the volume, we can
assert that large moduli in this case support the decompactification
scenarios. Looking at the right hand side of the previous inequality, this
construction has similarities with a compact direction in KK interpretation where the  DC  can be expressed as follows 
\begin{equation}
R_{KK}(\phi _{1})\sim R_{KK}(\phi _{2})e^{\alpha \Delta \phi }  \label{Eq3}
\end{equation}
where $\alpha$ is a free parameter. Thanks to the coexistence curve of the large and small black hole phases and
the WGC, we could  approach  the DC   statement 
in terms of the size of the two phases. Specifically, by requiring that
small phases adhere to the NoGSC, we explicitly require that the moduli
distance should  be small as 
\begin{equation}
\Delta \phi <<\mathcal{O}(1)  \label{Eq4}
\end{equation}%
In this picture, the small black hole phase, characterized by $r_{0}\sim
R_{0}e^{a\phi /2}$, naturally defines a particle-like excitation in the
presence of a compact extra dimension $R_{KK}$. This motivates an
interpretation of the compact direction as the  DD  $R_{D}$ whose
size is controlled by the modulus $\phi $. With respect to  Eq. (\ref{Eq2}) and  Eq.(%
\ref{Eq3}), such a  DD  can be expressed as
\begin{equation}
R_{D}\left( \phi \right) \sim R_{0}e^{a \phi }.  \label{5}
\end{equation}%
Along this dimension, the lightest states (the lowest KK modes)  describing  the small black hole phase are naturally long-lived and weakly coupled.
The corresponding mass reads as
\begin{equation}
m_{DM}\left( \phi \right) \sim R_{D}\left( \phi \right) ^{-1}\sim
R_{0}^{-1}e^{-a \phi }.  \label{6}
\end{equation}%
According to the DC  set in Eq. (\ref{Eq3}) with the required smallness
moduli excursions (\ref{Eq4}) where $m_{DM}\left( \phi _{1}\right) \sim
m_{DM}\left( \phi _{2}\right) e^{-a \Delta \phi }$ , these  DM 
states remain stable as

\begin{equation}
m_{DM}\sim \frac{2^{1/4}}{\sqrt{M\ell _{p}}}e^{-a \phi }.  \label{7}
\end{equation}%
Thus, this makes them a promising  DM  candidate. Indeed, since the
detection prospects of such a DM candidate is governed by the the modulus $%
\phi $ under the regime  relying on Eq. (\ref{Eq4}).  In fact,  its interaction strength weighted by
the corresponding coupling $g_{DM}\sim e^{-a \phi }\simeq \left(
1-a \phi \right) $ $\sim \mathcal{O}(1)$ is unsuppressed, leading to a
potentially observable interaction rate \cite{100,101}. This links the
microscopic properties of  the black holes and the  extra dimensional physics to
observable phenomena in the dark sector.

\section{ Conclusions}

Inspired by the idea that the cosmological constant could be considered as a
time dependent quantity, we have provided a scenario bridging certain
swampland conjectures from a new looking at thermodynamic black holes.  Precisely, we  have addressed phase transitions by proposing a new approach to the thermodynamics of charged black holes.
Interpreting the radial metric function at the horizon  as a state equation, we have
discussed certain  structure transitions and coexistence curves. Considering an
effective potential with a quasi static scalar field, we have elaborated an
interplay between certain swampland conjectures.  By examining the implications of the WGC, the Gibbs free energy appears to be able to distinguish between light and heavy phases, which leads to direct statements in the form of the DC.  Supported by KK
interpretations of scalar fields that could be incorporated into string theory, we have  included   DD  and  DM contributions    in such thermodynamic insights of charged black holes.

This work leaves  certain open  questions. One possible approach would be to explore other aspects of this issue, particularly rotation  behaviors.  This could add extra   aspects    offering  a jump to the optical  aspect involving shadow curves needed to make contact with empirical investigations including EHT collaborations.   Other transitions  such as  Hawking-Page ones could be incorporated in such activities.  Alternatively, it would be also  interesting to explore connections with de Sitter thermodynamics in black hole physics to incorporate further conjectures. With regard to the gauging of global symmetries, we expect that the discontinuity of the partial derivative of the Gibbs free energy with respect to the bundles connection measure suggests the fundamental role that the free energy might play in the detection of UV completions. Interestingly, by considering black holes evaporating down to Planck scales, the the allowed small phase size becomes smaller. Consequently, this could relate the DM mass to primordial black holes. At first sight,  this  deserves further study. We hope to address these questions in upcoming works.

\end{document}